\documentclass{PoS}

\usepackage{subfigure}

\title{Z(N) dependence of the pure Yang-Mills gluon propagator in the Landau gauge near Tc}

\ShortTitle{Z(N) dependence of the Landau gauge gluon propagator near Tc}

\author{Orlando Oliveira, \speaker{Paulo J. Silva} \\
        Center for Computational Physics, University of Coimbra\\
        E-mail: \email{orlando@teor.fis.uc.pt}, \email{psilva@teor.fis.uc.pt}}

\abstract{The Z(N) dependence of the pure Yang-Mills gluon propagator, in the Landau gauge, is investigated at finite temperature for N=3. Special attention will be given to the behaviour near the critical temperature $T_c$. Our simulations show a complex pattern as expected in a first order phase transition. Furthermore, we identify an order parameter directly associated with the breaking of the SU(3) center symmetry.}

\FullConference{The 32nd International Symposium on Lattice Field Theory\\
		 23-28 June, 2014\\
		 Columbia University New York, NY}

\begin{document}

\section{Introduction and motivation}

At low temperatures quarks and gluons are confined within color singlet states but for sufficiently high temperatures they become deconfined and
behave as essentially free particles. The order parameter for the confinement-deconfinement QCD transition is the Polyakov loop
\begin{equation}
L = \langle L( \vec{x} ) \rangle \, \propto \, e^{-F_q/T} \ ,
\end{equation}
where $F_q$ is the quark free energy. On the lattice, the Polyakov loop is given by
\begin{displaymath}
L( \vec{x} ) = \mathrm{Tr} \prod^{N_t-1}_{t=0} \, U_4(\vec{x},t) \ .
\end{displaymath}
For temperatures below the critical temperature $T_c \sim 270$ MeV (for the pure gauge theory) the spacetime average Polyakov loop takes the
value $L=0$, i.e. $F_q = + \infty$, whereas for $T>T_c$ $L \neq 0$, corresponding to a $F_q = 0$. Finite volume effects make the transition at 
$ T =T_c$ smoother and $L$ has a sharp transition from a small nonvanishing value to a number just below one and that approaches one from below
as $T$ is increased.

On the lattice, the Wilson gauge action is invariant under a center transformation where the temporal links on a hyperplane $x_4=const$ 
are multiplied by an element of the center group $z \in Z_3 =  \{e^{- i 2 \pi/3}, 1, e^{ i 2 \pi/3} \}$. In what concerns center transformations,
the Polyakov loop is changed according to $L(\vec{x}) \rightarrow z \, L(\vec{x})$. By definition, the Polyakov loop is a gauge invariant
quantity and, therefore, a center transformation is not a gauge transformation. Center transformations connect gauge configurations
which have exactly the same action and contribute equally to the QCD generating functional, but do not belong to the same gauge orbit, 
i.e. they are not connected by gauge transformations. 

The phase of the Polyakov loop can be used to characterize different regions of the SU(3) manifold of the gauge configurations. As shown in, for example, \cite{polyloop} for $T<T_c$ the  local $L( \vec{x} )$ phase is equally distributed among the 
possible values and it follows that the average of the Polyakov loop over all lattice sites is $L  \approx 0$ (center symmetric phase).  
On the other hand, for $T>T_c$ the various possible phases are not equally populated and $L \neq 0$ (spontaneous broken center symmetric phase).

The gluon propagator is a fundamental non gauge invariant QCD correlation function that, for example, can be used to define a
potential for heavy quarkonium. Our aim is to investigate how the Landau gauge gluon propagator changes with $T$ near the critical temperature $T_c$ and how it correlates with
the phase of the Polyakov loop. In the present report we will focus only on pure gauge sector and ignore possible contributions from quarks,
which break explicitly the center symmetry.

\section{Lattice setup}

At finite temperature, the Landau gauge gluon propagator has two independent form factors,
\begin{equation}
D^{ab}_{\mu\nu}(\hat{q})=\delta^{ab}\left(P^{T}_{\mu\nu} D_{T}(q_4,\vec{q})+P^{L}_{\mu\nu} D_{L}(q_4,\vec{q}) \right)  \ .
\end{equation}
It is known that the electric $D_L$ and magnetic $D_T$ form factors change with $T$, with $D_L$ changing more dramatically than
$D_T$ -- see e.g. \cite{gluonmass, Aouane, cucc} and references therein. For the present work, the two form factors were computed on
lattices whose physical volume is about $\sim (6.5 \textrm{fm})^3$; we have considered coarser lattices, with $a \sim 0.12~ \textrm{fm}$, and finer lattices,
with $a \sim 0.09 ~ \textrm{fm}$. The lattice setup is described in tables \ref{coarsesetup} and \ref{finesetup}.
All results reported for $D_L$ and $D_T$ refer to 100 configurations per ensemble.

\begin{table}
\begin{minipage}[b]{0.45\linewidth}
\centering
\begin{tabular}{ll@{\hspace{0.5cm}}l@{\hspace{0.5cm}}l@{\hspace{0.5cm}}l}
\hline
Temp.   & $L^3_s  \times L_t$ & $\beta$ & $a$  & $L_s a$ \\
 (MeV)  &                                 &              & (fm)  & (fm) \\
\hline
 265.9  & $54^3 \times 6$   &   5.890   &  0.1237  & 6.68 \\
 266.4  & $54^3 \times 6$   &   5.891   &  0.1235  & 6.67 \\
 266.9  & $54^3 \times 6$   &   5.892   &  0.1232  & 6.65  \\
 267.4  & $54^3 \times 6$   &   5.893   & 0.1230  & 6.64  \\
 268.0  & $54^3 \times 6$   &   5.8941  & 0.1227  & 6.63 \\
 268.5  & $54^3 \times 6$   &   5.895   & 0.1225  & 6.62  \\
 269.0  & $54^3 \times 6$   &   5.896   & 0.1223  & 6.60  \\
 269.5  & $54^3 \times 6$   &   5.897   & 0.1220  & 6.59 \\
 270.0  & $54^3 \times 6$   &   5.898   & 0.1218  & 6.58 \\
 271.0  & $54^3 \times 6$   &   5.900   & 0.1213  & 6.55 \\
 272.1  & $54^3 \times 6$   &   5.902   & 0.1209  & 6.53 \\
 273.1  & $54^3 \times 6$   &   5.904   & 0.1204   & 6.50 \\
\hline
\end{tabular}
\caption{Simulation setup: coarse lattices.}
\label{coarsesetup}
\end{minipage}
\hspace*{0.05\linewidth}
\begin{minipage}[b]{0.45\linewidth}
\centering
\vspace*{-0.5cm}
\begin{tabular}{ll@{\hspace{0.5cm}}l@{\hspace{0.5cm}}l@{\hspace{0.5cm}}l}
\hline
Temp.   & $L^3_s  \times L_t$ & $\beta$ & $a$  & $L_s a$ \\
 (MeV)  &                                 &              & (fm)  & (fm) \\
\hline
 269.2  & $72^3 \times 8$   &   6.056    & 0.09163 & 6.60  \\
 270.1  & $72^3 \times 8$   &   6.058    & 0.09132 & 6.58  \\
 271.0  & $72^3 \times 8$   &   6.060    & 0.09101 & 6.55  \\
 271.5  & $72^3 \times 8$   &   6.061    & 0.09086 & 6.54 \\
 271.9  & $72^3 \times 8$   &   6.062    & 0.09071 & 6.53  \\
 272.4  & $72^3 \times 8$   &   6.063    & 0.09055 & 6.52  \\
 272.9  & $72^3 \times 8$   &   6.064    & 0.09040 & 6.51  \\
 273.3  & $72^3 \times 8$   &   6.065    & 0.09025 & 6.50  \\
 273.8  & $72^3 \times 8$   &   6.066    & 0.09010  & 6.49 \\
\hline
\end{tabular}
\caption{Simulation setup: fine lattices.}
\label{finesetup}
\end{minipage}
\end{table}

The SU(3) gauge configurations were generated using a combination of heat bath 
and overrelaxation updates and,  for each configuration, three
independent gauge fixings after the center transformation 
$U^\prime_4 ( \vec{x}, t = 0) = z \,U_4 ( \vec{x}, t = 0)$ 
were performed using all possible values $z \in Z_3$.
For each of the gauge fixed configurations the Polyakov loop 
$\langle L \rangle = |L| e^{i \theta} $ was computed and the 
configurations were classified according to
\begin{displaymath}
                           - \pi < \theta \le -\frac{\pi}{3}    \hspace{0.4cm} (\mbox{Sector -1}) , \hspace{1cm}
                            -\frac{ \pi}{3}  < \theta \le \frac{\pi}{3}   \hspace{0.4cm} (\mbox{Sector 0}) , \hspace{1cm}
                            \frac{ \pi}{3} < \theta \le \pi   \hspace{0.4cm} (\mbox{Sector 1}) \ .
\end{displaymath} 
For a given configuration, the values of $L( \vec{x} )$ are not clearly on top \cite{polyloop} of the possible phases of $Z(3)$ center symmetry.
Indeed, above $T_c$ the values of $\theta$, for each gauge fixed configuration, are typically distributed around $\theta = 0, \pm 2 \pi /3$.
In this preliminary study we do not investigate the effects associated with the introduction of cutoffs on the phase of the Polyakov loop to identify the
various sectors.

\begin{figure}
\vspace{0.2cm}
   \centering
   \subfigure[Sector -1]{ \includegraphics[angle=-90,scale=0.16]{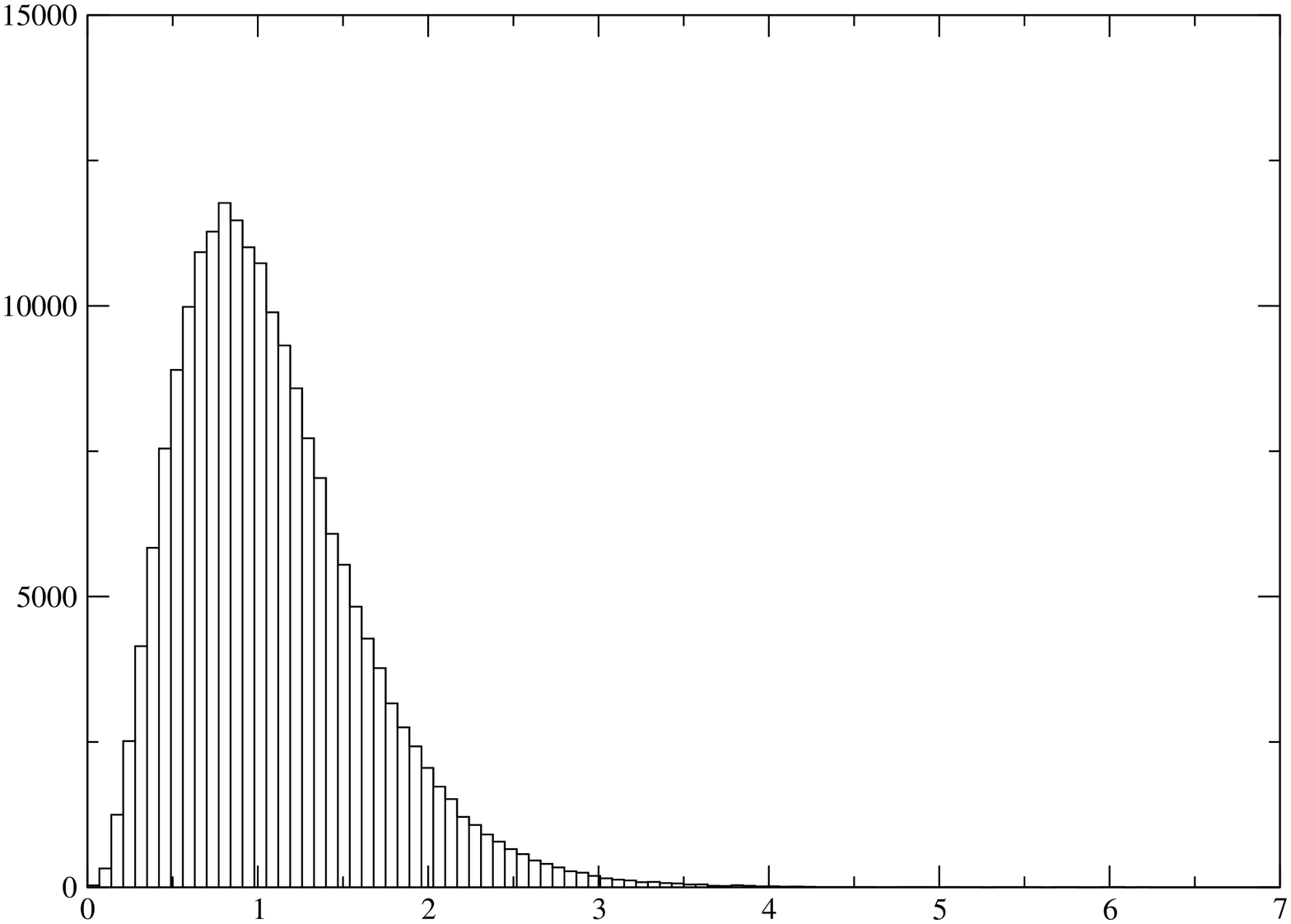} } \quad
   \subfigure[Sector 0]{ \includegraphics[angle=-90,scale=0.16]{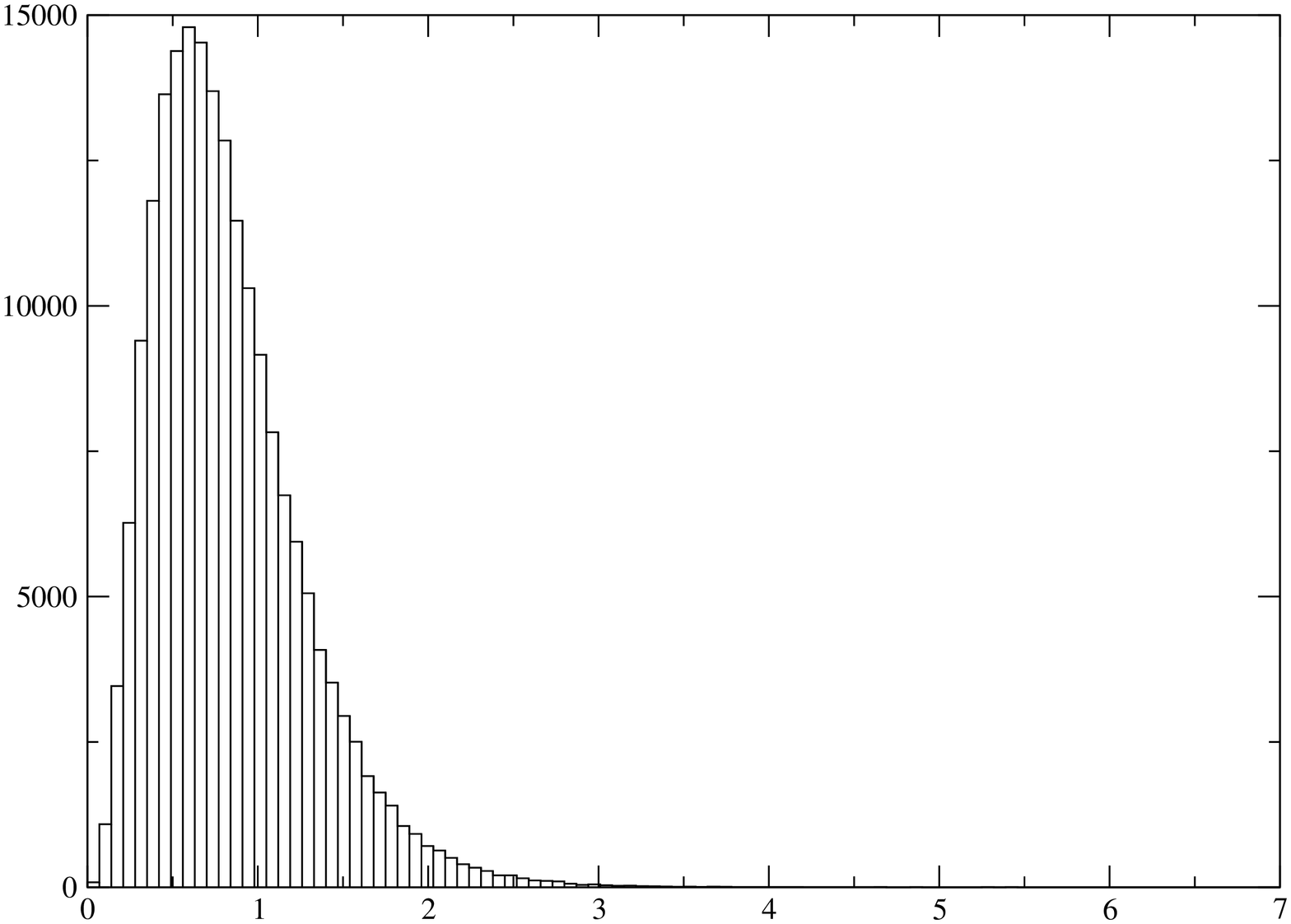} } \quad
\subfigure[Sector 1]{ \includegraphics[angle=-90,scale=0.16]{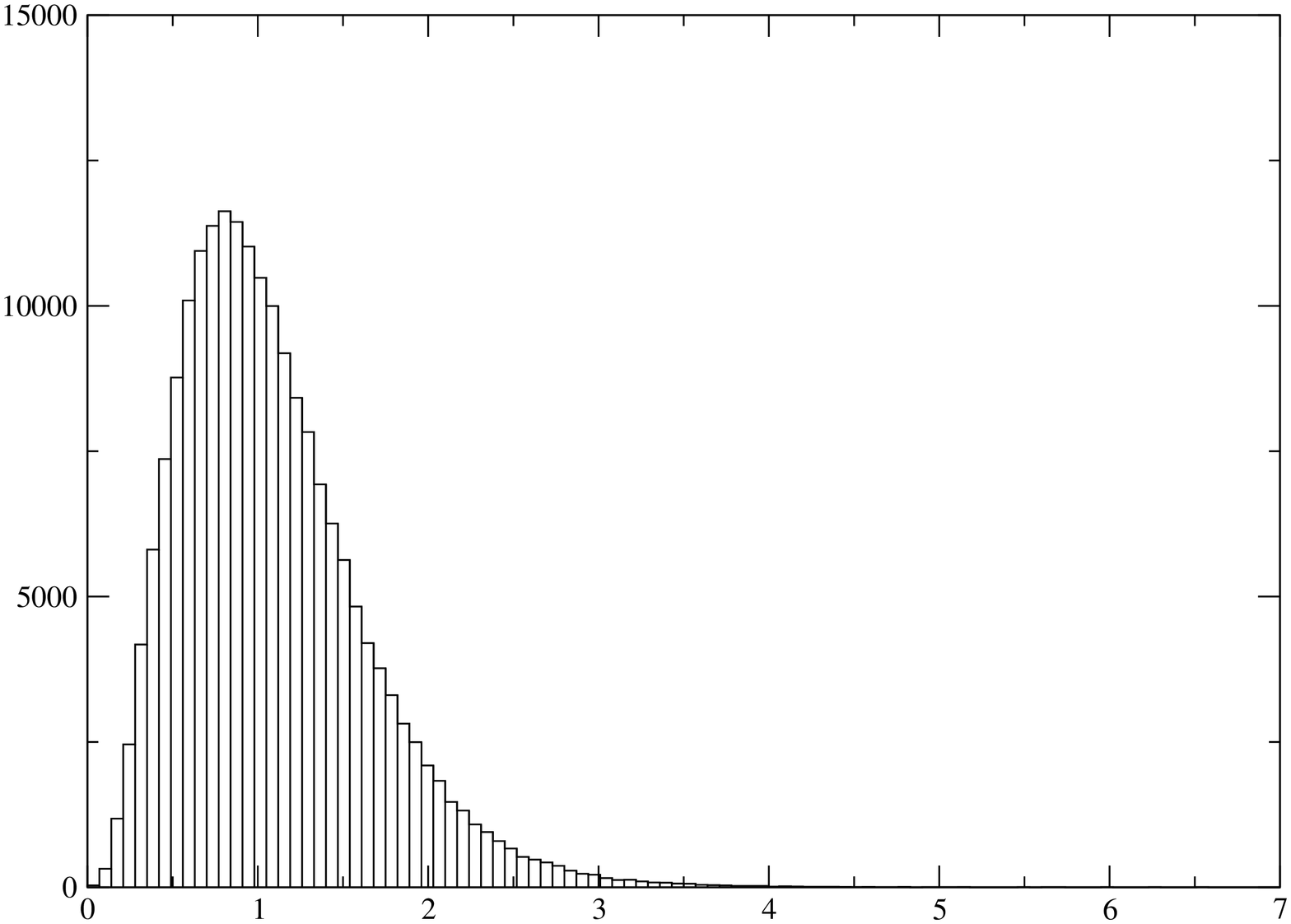} }
  \caption{Histograms exhibiting the distribution of the distance between the temporal links and the unit matrix; we use a $32^3\times6$, $\beta=6.0$ configuration with $T = 324$ MeV. }
   \label{histF}
\end{figure}

For the computation of the gluon field and, therefore, the gluon propagator 
we rely on the usual definitions that can be found in
e.g.~\cite{gluonmass,volume}. 
Naively, one could claim that, in what concerns the definition 
of the gluon field for sectors $\pm 1$, one should 
subtract a constant term associated with the phase of the Polyakov loop. 
However, when going to the momentum space, such a subtraction only changes the
zero momentum gluon field leaving all the other momenta unchanged. 
On the other hand, some authors (e.g. \cite{rank95}) claim that in sectors other than the zero sector, the links are not close to the unit matrix, and therefore the usual formula to compute the gluon field is not valid. 
In figure \ref{histF} we report the distribution of the distance 
(as defined in \cite{zwanziger}) of the temporal links to the unit matrix for a configuration in the confined phase;  the difference between the various plots does not support a different definition for $A_{\mu}(x)$ in the different sectors. 
We will report elsewhere~\cite{futuro} the discussion on the connection between the lattice link variable, the gluon field and the gluon propagator
for the various sectors.

In order to reduce lattice artifacts, we have performed a conic cut~\cite{Aouane} for momenta above 1 GeV and take into account all lattice data 
below 1 GeV. The propagators described below refer to renormalized data chosen such that
\begin{displaymath}
D_{L,T}(\mu^2) = Z_R \,  D^{Lat}_{L,T} (\mu^2) = 1 /\mu^2 
\end{displaymath}
for a renormalization scale of $\mu = 4$ GeV. The form factors $D_L$ and $D_T$ were renormalized independently. 
The simulations show renormalization constants that are compatible within one standard deviation for each of the form factors and 
between the different $Z(3)$ sectors.

\section{Gluon Propagator near $T_c$}

At finite temperature, the two form factors associated to the gluon propagator have been computed several times and their dependence with $T$ 
has been studied --- see, for example, \cite{gluonmass, Aouane, cucc} and references therein. Typically, the computation is performed either not taken into account which sector of the SU(3) manifold the configurations belong
or projecting into the zero sector.

\begin{figure}[h] 
\vspace{0.2cm}
   \centering
   \subfigure[Longitudinal component.]{ \includegraphics[scale=0.24, angle=-90]{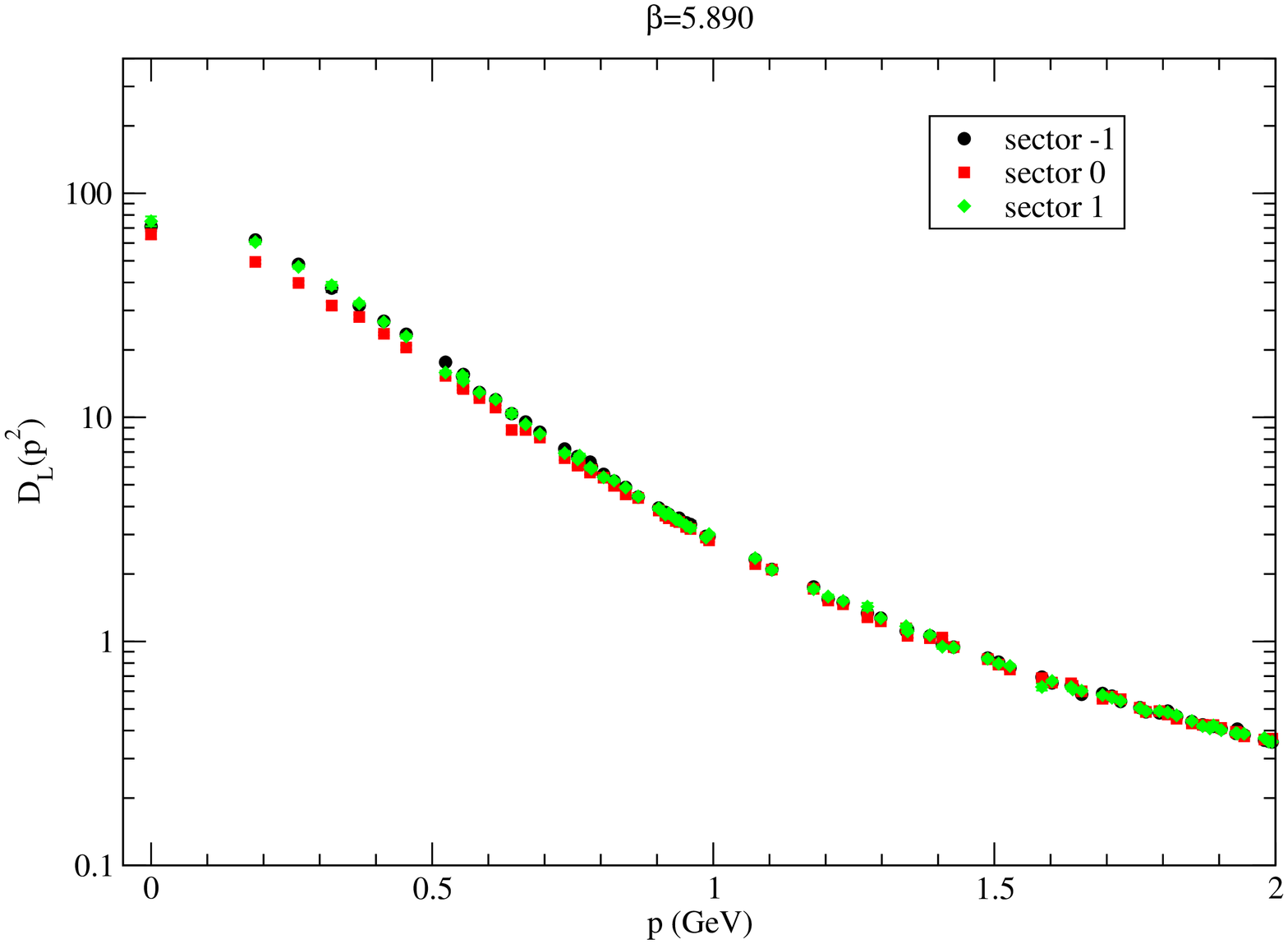} } \qquad
   \subfigure[Transverse component.]{ \includegraphics[scale=0.24, angle=-90]{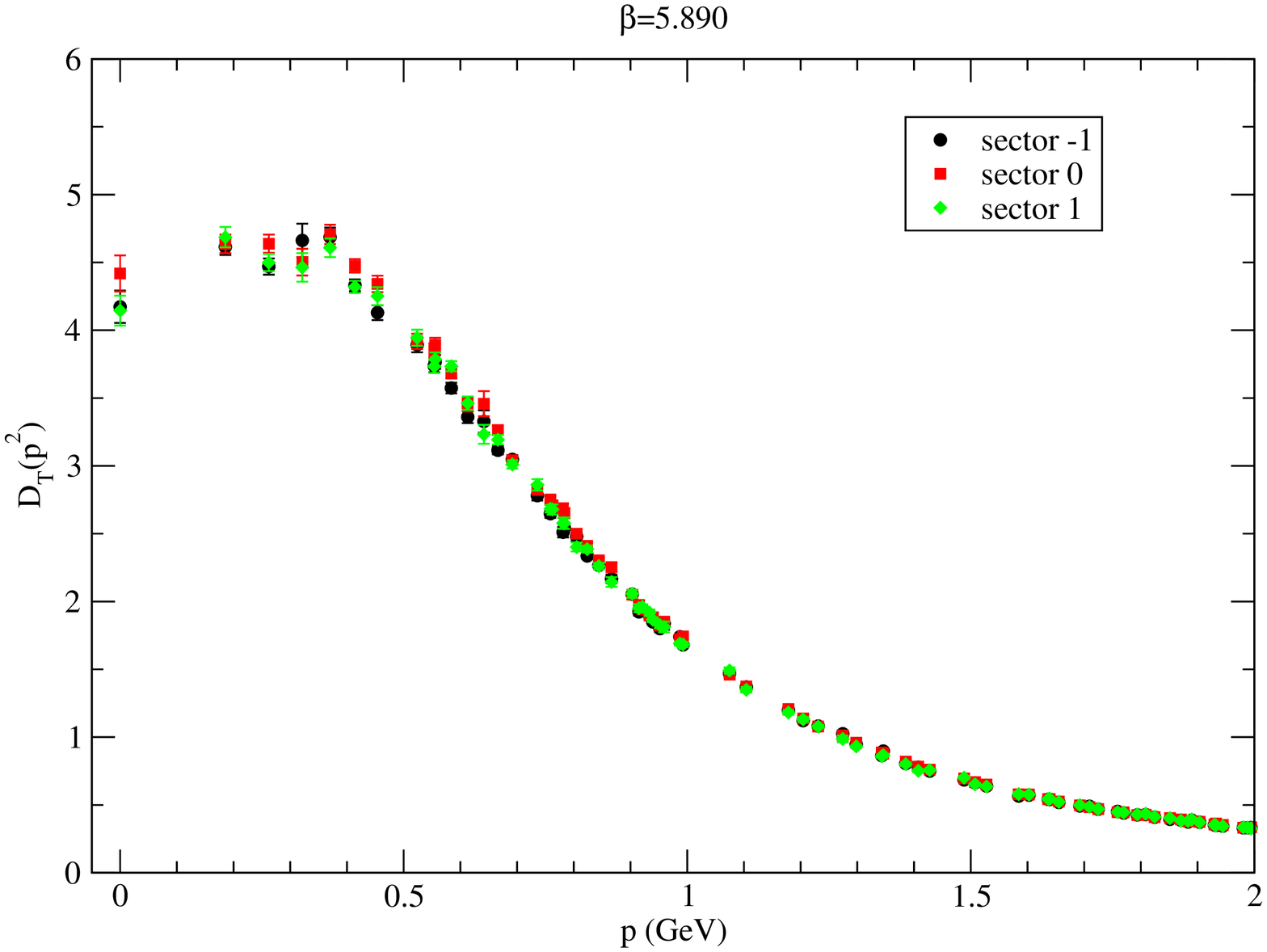} }
  \caption{Coarse lattices, below $T_c$.}
   \label{coarse-cold}
\end{figure}

\begin{figure}[h] 
\vspace{0.2cm}
   \centering
   \subfigure[Longitudinal component.]{ \includegraphics[scale=0.24, angle=-90]{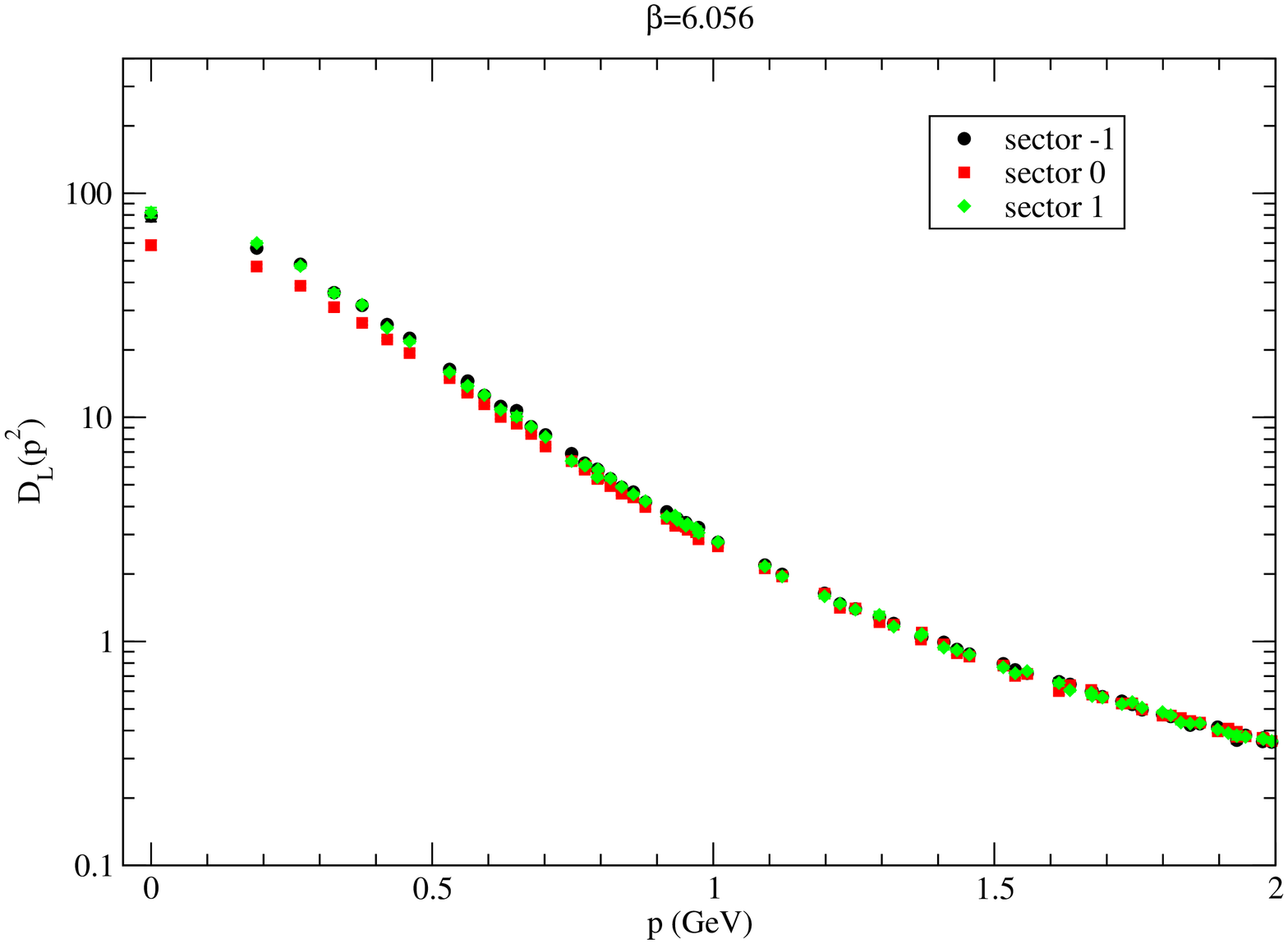} } \qquad
   \subfigure[Transverse component.]{ \includegraphics[scale=0.24, angle=-90]{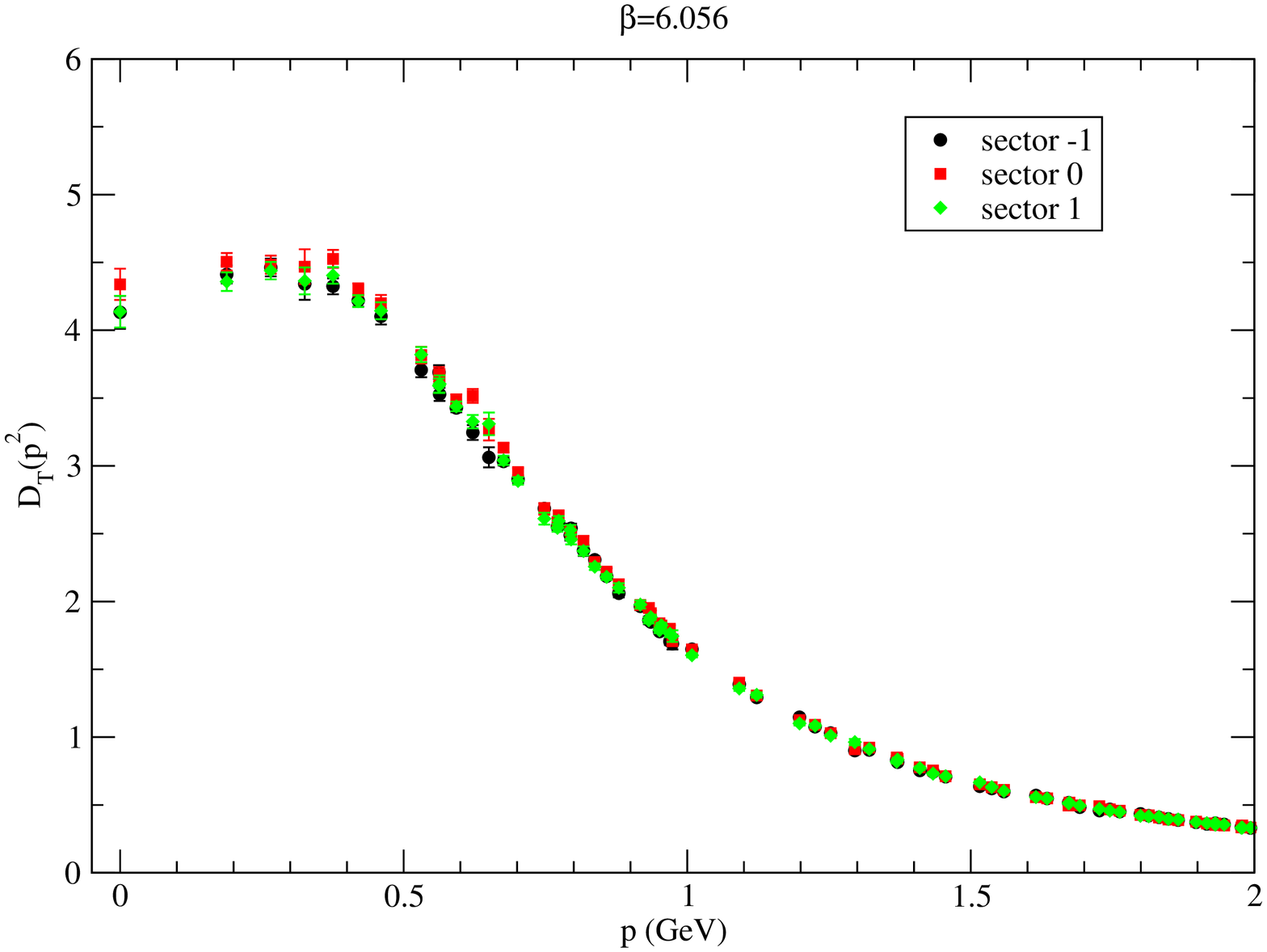} }
  \caption{Fine lattices, below $T_c$.}
   \label{fine-cold}
\end{figure}

\begin{figure}[h] 
\vspace{0.2cm}
   \centering
   \subfigure[Longitudinal component.]{ \includegraphics[scale=0.24, angle=-90]{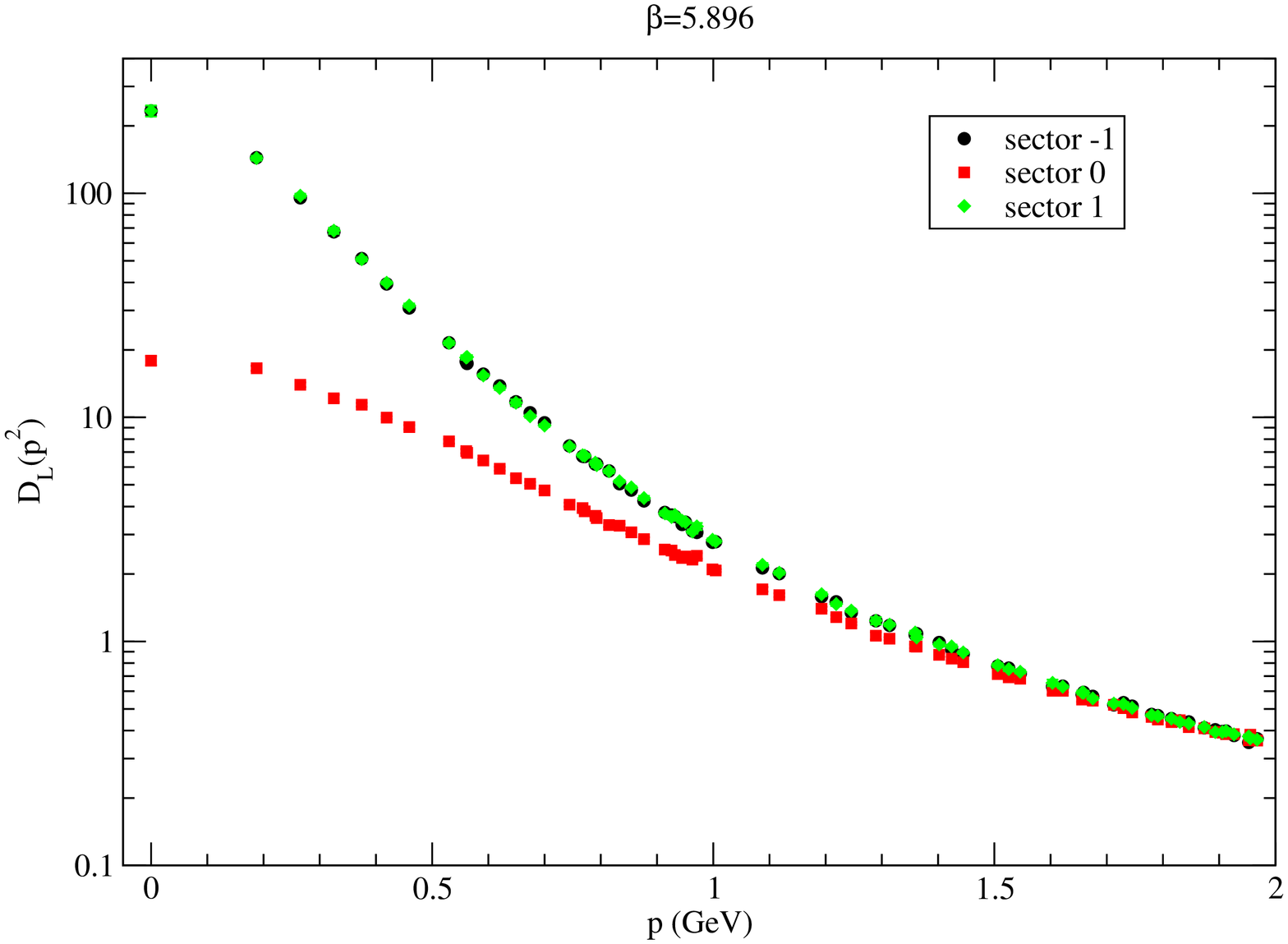} } \qquad
   \subfigure[Transverse component.]{ \includegraphics[scale=0.24, angle=-90]{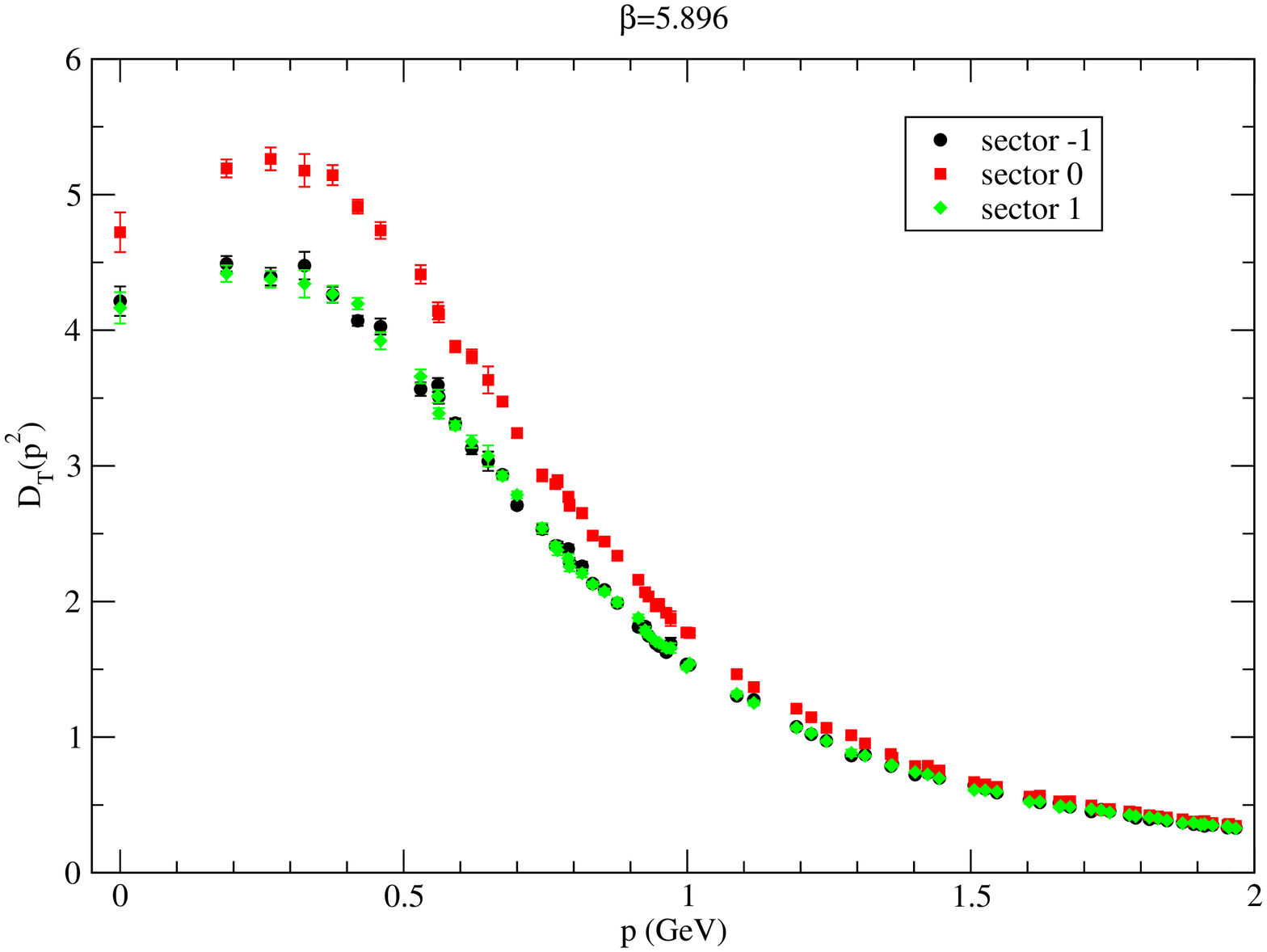} }
  \caption{Coarse lattices, above $T_c$.}
   \label{coarse-hot}
\end{figure}

\begin{figure}[h] 
\vspace{0.2cm}
   \centering
   \subfigure[Longitudinal component.]{ \includegraphics[scale=0.24, angle=-90]{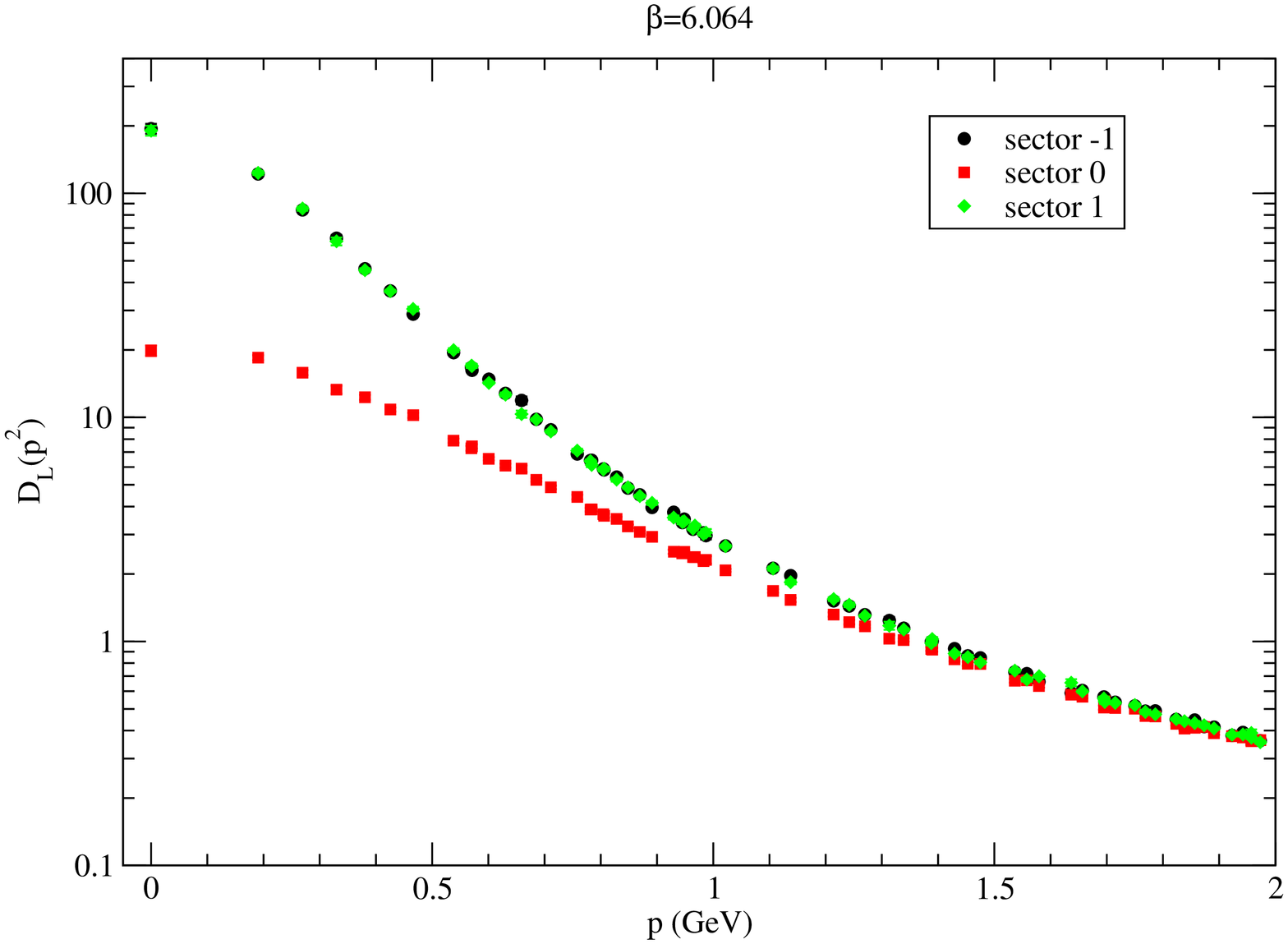} } \qquad
   \subfigure[Transverse component.]{ \includegraphics[scale=0.24, angle=-90]{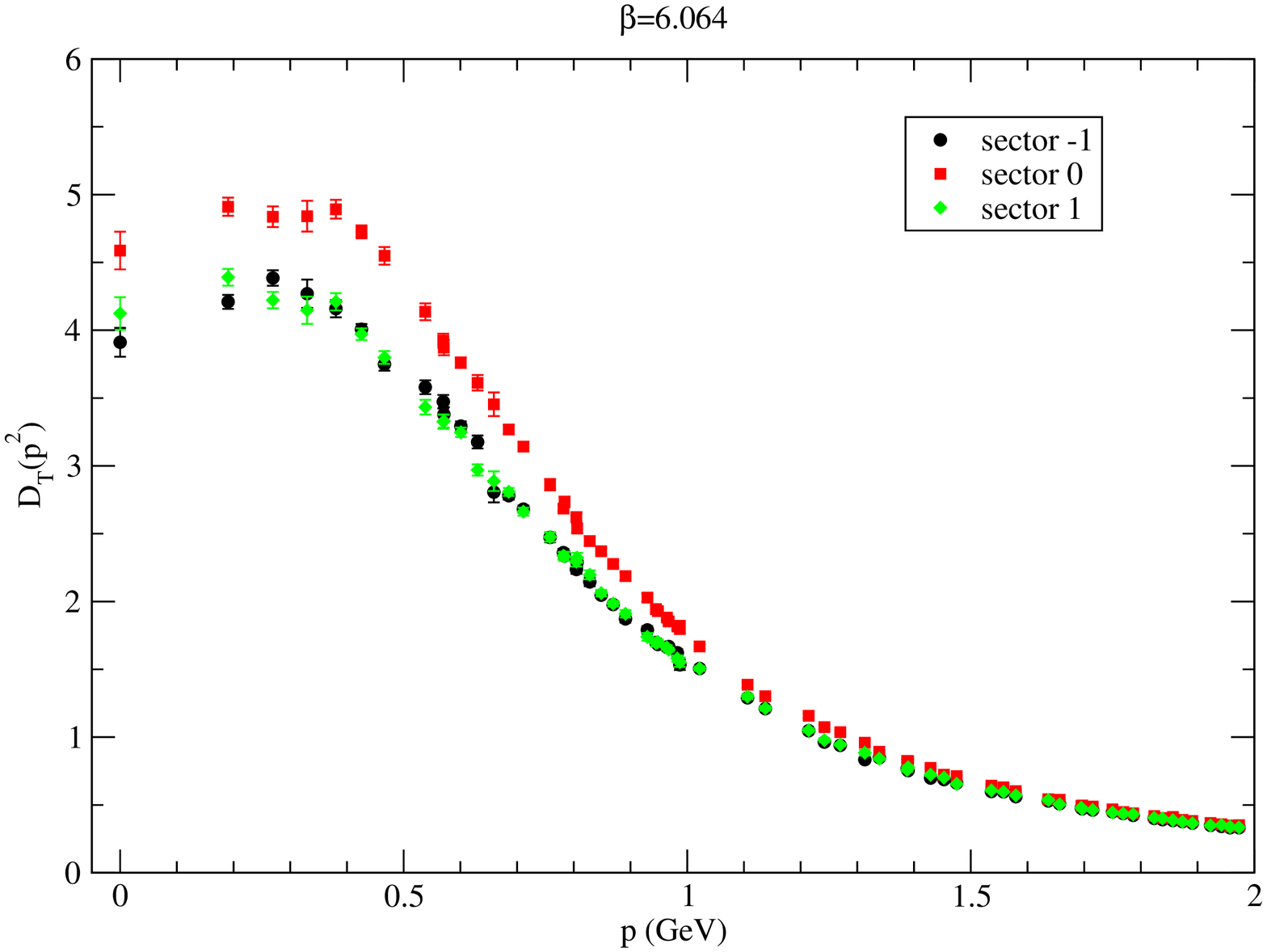} }
  \caption{Fine lattices, above $T_c$.}
   \label{fine-hot}
\end{figure}

Figures \ref{coarse-cold}, \ref{fine-cold}, \ref{coarse-hot} and \ref{fine-hot} show the propagators below and above $T_c$ for the various sectors.
For $T < T_c$, there is a slight enhancement of the longitudinal propagator in the $\pm1$ sectors relative to the $0$ sector. 
On the other hand, the transverse propagator in $\pm1$ sectors is slightly supressed. Above the deconfinement transition, it is observed
a huge enhancement of the electric form factor and a sizeable suppression of the magnetic component in the $\pm1$ sectors, relative to the $0$ sector. Clearly,
above $T_c$ the propagators in each sector have different functional forms, suggesting that the dynamics associated with the configurations
in each of the sectors of the SU(3) manifold characterized by the phase of the Polyakov loop is also different. Furthermore, the results seem to
suggest that one can use the difference between the propagators in the various sectors to distinguish if a given configuration is either on
the confined or deconfined phase.

\begin{figure}[h] 
\vspace{0.2cm}
   \centering
   \subfigure[$\beta=6.060$]{ \includegraphics[scale=0.24, angle=-90]{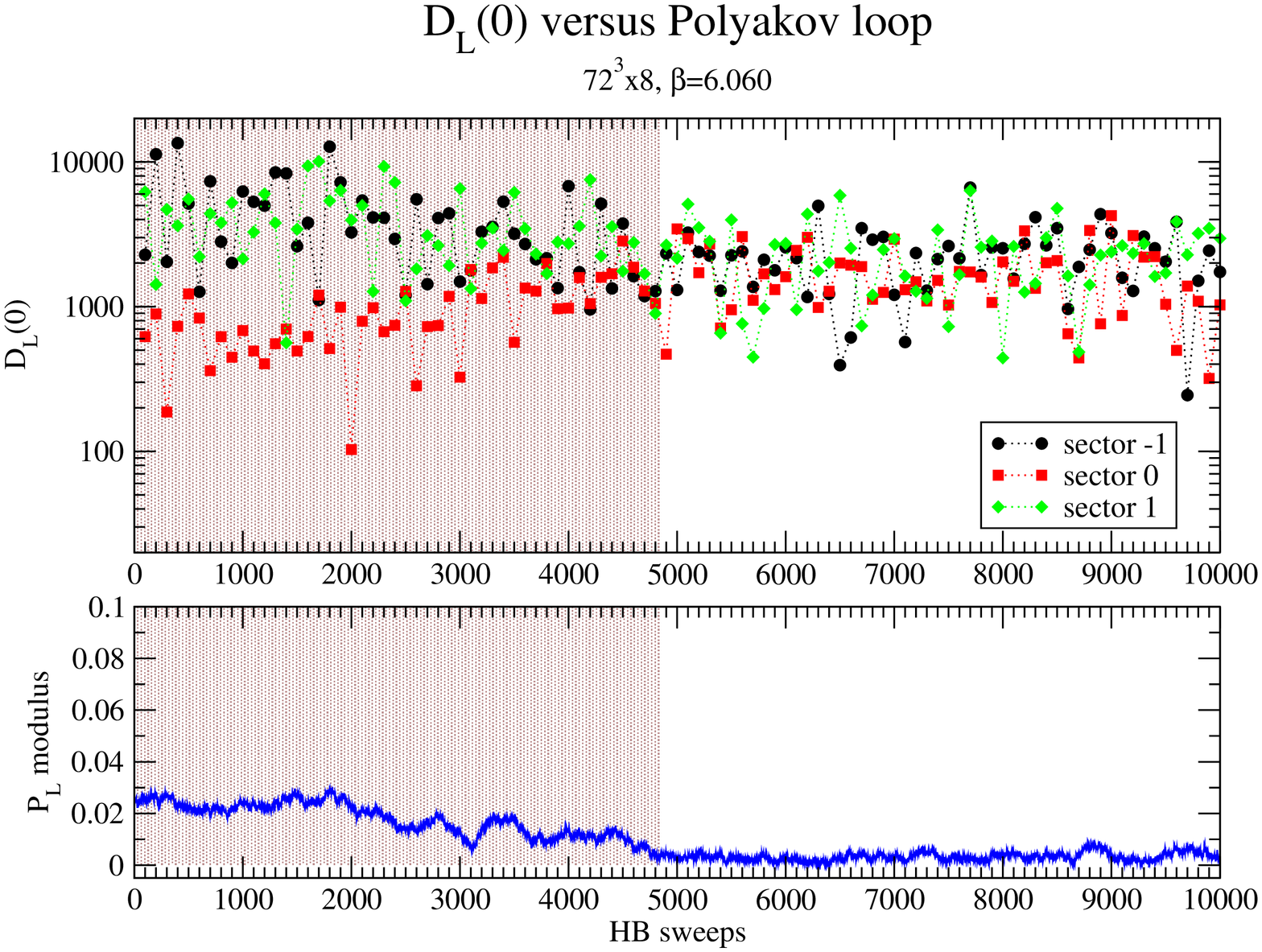} } \qquad
   \subfigure[$\beta=6.066$]{ \includegraphics[scale=0.24, angle=-90]{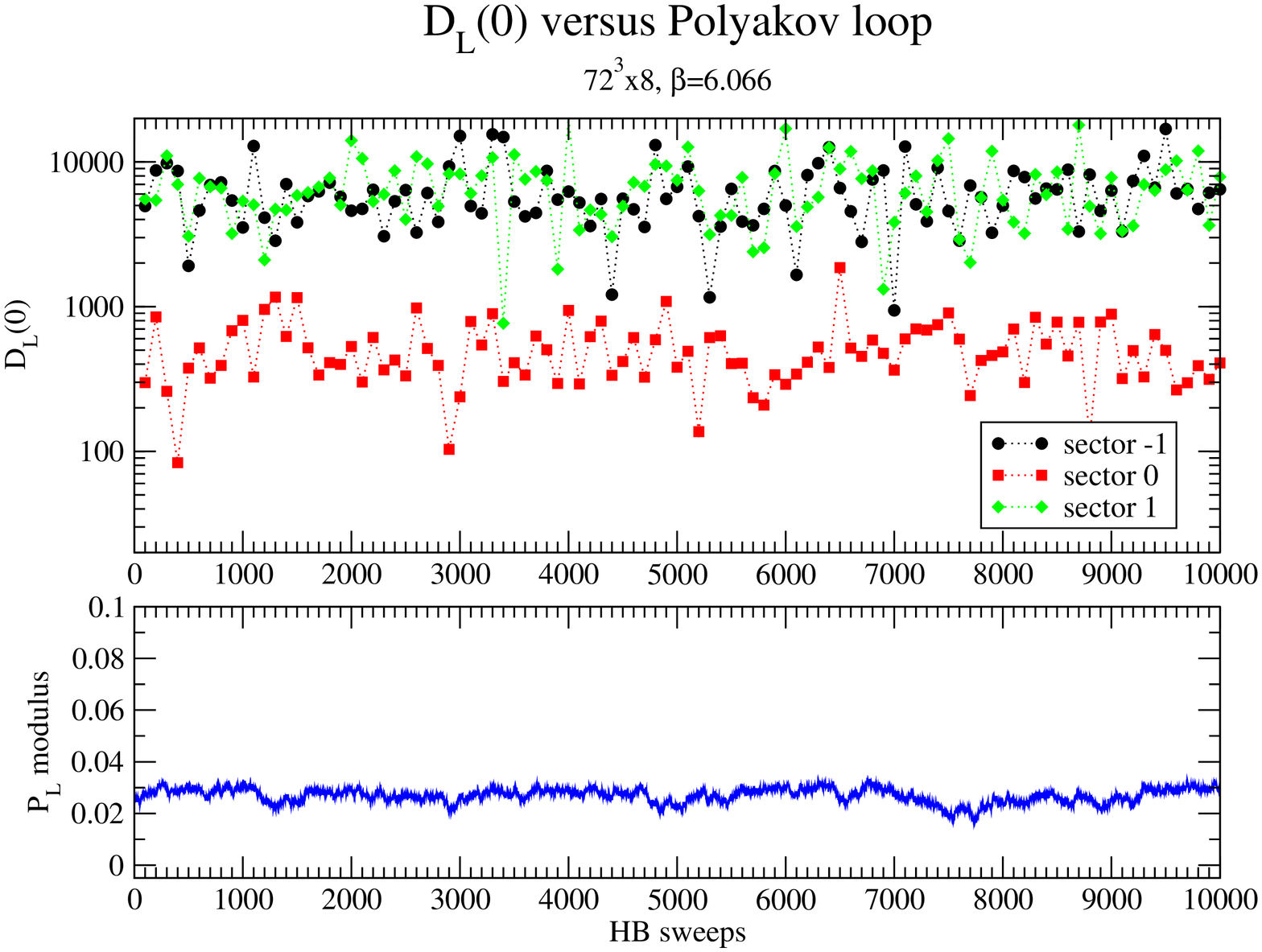} } 
  \caption{Unrenormalized Polyakov loop sampling history for $T = 271.0$ MeV  and $T = 273.8.0$ MeV  (fine lattices).}
   \label{PLhist}
\end{figure}

\begin{figure}[h] 
\vspace{0.5cm}
\centering
 \includegraphics[scale=0.35]{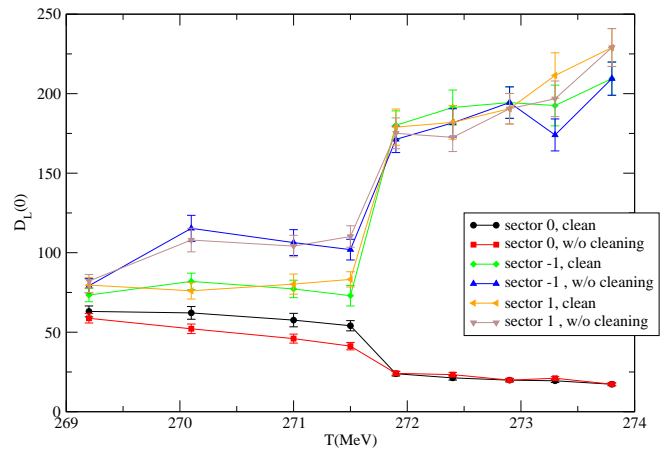}
  \caption{$D_L(0)$ as a function of T (fine lattices) near the phase transition.}
   \label{DLzeroT}
\end{figure}

In order to test this hypothesis, one can look at the Polyakov loop sampling history for temperatures around $T_c$. Figure
\ref{PLhist} illustrates the correlation between the modulus of the Polyakov loop and $D_L (0)$ measured in
each sector of the configuration manifold for $T = 270.1$ and $273.8$ MeV, respectively. As observed, when $|L|$ becomes smaller
(confined phase), $D_L(0)$ has a unique value for all sectors. On the other hand, when $|L|$ becomes larger, $D_L(0)$ in sectors
$\pm 1$ decouple from the $0$ sector values. Finally, in figure \ref{DLzeroT} we show $D_L(0)$ as a function of $T$ near the phase transition.

Figures \ref{PLhist} and \ref{DLzeroT} support the idea that the separation between the propagators computed in different sectors provide an indication if the configuration is in the confined or deconfined phase.

In some cases, the simulation mixes both the confined and deconfined phase; one example can be seen in the left plot of Fig. \ref{PLhist}. For such cases, it is a sensible approach to clean the ensemble by removing the configurations in the wrong phase (marked using a shadow in the graph). In Fig. \ref{DLzeroT} we compare the dependence of $D_L(0)$ with $T$ with and without cleaning. We observe that the discontinuity in $D_L(0)$ gets stronger for clean ensembles. Certainly, this is an issue which can change the conclusions reported recently in \cite{cucc} about the nature of the transition in the longitudinal propagator at the critical temperature.

\section*{Acknowledgments}

Simulations have been carried out in Milipeia
and Centaurus computer clusters at the University of
Coimbra, with the help of Chroma \cite{chroma} and PFFT \cite{pfft} 
libraries. Paulo Silva acknowledges support 
by FCT under contract SFRH/BPD/40998/2007. Further support 
has been provided by projects CERN/FP/123612/2011, 
CERN/FP/123620/2011 and PTDC/\-FIS/100968/2008, developed 
under initiative QREN financed by UE/FEDER through 
Programme COMPETE.

\end{document}